\documentclass[aps,english,prb,twocolumn,showpacs,superscriptaddress]{revtex4}
\usepackage{graphicx}

\begin{document}

\title{Self-assembly mechanisms of short atomic chains on single layer graphene and boron nitride}

\author{V. Ongun \" Oz\c celik}
\affiliation{UNAM-National Nanotechnology Research Center, Bilkent University, 06800 Ankara, Turkey}
\affiliation{Institute of Materials Science and Nanotechnology, Bilkent University, Ankara 06800, Turkey}

\author{S. Ciraci}\email{ciraci@fen.bilkent.edu.tr}
\affiliation{UNAM-National Nanotechnology Research Center, Bilkent University, 06800 Ankara, Turkey}
\affiliation{Institute of Materials Science and Nanotechnology, Bilkent University, Ankara 06800, Turkey}
\affiliation{Department of Physics, Bilkent University, Ankara 06800, Turkey}

\begin{abstract}
Nucleation and growth mechanisms of short chains of carbon atoms on single-layer, hexagonal boron nitride (h-BN), and short BN chains on graphene are investigated using first-principles plane wave calculations. Our analysis starts with the adsorption of a single carbon ad-atom and examines its migrations. Once a C$_2$ nucleates on h-BN, the insertion of each additional carbon at its close proximity causes a short segment of carbon atomic chain to grow by one atom at at a time in a quaint way: The existing chain leaves its initial position and subsequently is attached from its bottom end to the top of the carbon ad-atom. The electronic, magnetic and structural properties of these chains vertically adsorbed to h-BN depend on the number of carbon atoms in the chain, such that they exhibit an even-odd disparity. An individual carbon chain can also modify the electronic structure with localized states in the wide band gap of h-BN. As a reverse situation we examined the growth of short BN atomic chains on graphene, which attribute diverse properties depending on whether B or N is the atom bound to the substrate. These  results together with ab-initio molecular dynamics simulations of the growth process reveal the interesting self-assembly behavior  of the grown chains. Furthermore, we find that these atomic chains enhance the chemical activity of h-BN and graphene sheets by creating active sites for the bonding of various ad-atoms and can act as pillars between two and multiple sheets of these honeycomb structures leaving wider spacing between them to achieve high capacity storage of specific molecules.
\end{abstract}

 \maketitle

\section{Introduction}

\begin{figure}
\includegraphics [width=8cm] {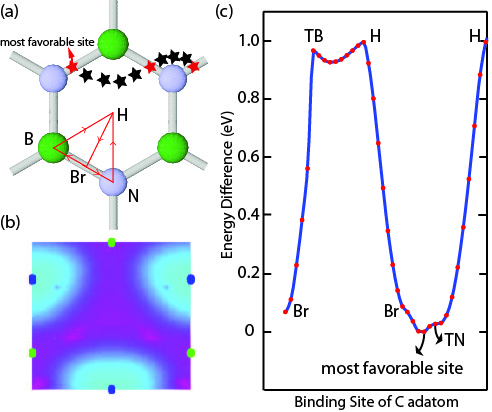}
\caption{Energy variation of single carbon atom adsorbed on various sites of single layer, 2D hexagonal boron nitride structure (h-BN) calculated in $(4\times4)$ supercell. (a) h-BN honeycomb structure on which the adsorption energies are calculated. Nitrogen and boron atoms are represented by blue and green balls, respectively. The most favorable binding site of carbon ad-atom is marked by the red star in the figure. The path of diffusion of carbon ad-atom with the minimum energy barrier of $\sim0.65eV$ is indicated by stars. (b) Complete energy landscape of C ad-atom on h-BN structure. Light blue regions show favorable sites and the energy barrier further increases as the color goes to dark blue and purple. (c) Energy variation of carbon ad-atom is shown along the path indicated by red arrows in (a). The energy difference between the most favorable site (indicated by red star) and the bridge(Br), top of boron(TB), hollow(H), top of nitrogen(TN) sites are calculated as $0.07 eV$, $0.95eV$, $1.00eV$ and $0.03eV$, respectively.}
\label{fig1}
\end{figure}

After the synthesis of graphene,\cite{novoselov2004, geim2007} a monolayer of $sp^2$-bonded honeycomb structure of carbon atoms, interest has focused on two dimensional nanostructures having honeycomb structures. Planar hexagonal boron nitride(h-BN) is an ionic honeycomb structure consisting of alternatively bonded boron and nitrogen atoms and is an analog of graphene.\cite{paszkowicz2002, liu2011} Despite the structural similarity, h-BN differs from graphene with its wide band gap and dielectric properties.\cite{watanabe2004} Various boron nitride structures like nanosheets,\cite{pacile2008} nanotubes\cite{chopra1995} and nanowires\cite{chen2006} have already been synthesized. Also, recent studies show that h-BN can be used to improve the current voltage properties of graphene transistors by improving the mobility of electrons in graphene as compared to graphene films on silicon substrates.\cite{dean2010} These properties hold promise for novel technological applications of h-BN structures.

A thorough understanding of the properties of h-BN structure and investigating its functionalization is important for future applications of this wide band gap material. Carbon, being in the same row of the periodic table with boron and nitrogen, is one of those foreign atoms which can greatly change the physical and chemical properties of h-BN. In a recent study,\cite{ataca2010} the effects of ad-atoms adsorbed on BN were investigated using first-principles calculations, and it was shown that high coverage of carbon ad-atoms can change the magnetic properties and band gap of the system, whereas ad-atoms induce localized states in the band gap at low coverage.

\begin{figure}
\includegraphics [width=8.5cm] {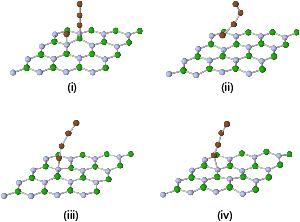}
\caption{Snapshots of the molecular dynamics simulation showing the formation of a short chain comprising four carbon atoms. The snapshots correspond to the initial, 20th, 40th and 120th steps of the molecular dynamics simulation done at 500K. Note that the formation of CAC(4) takes place as the CAC(3)leaves its initial bonding position and attaches to a single carbon ad-atom at close proximity from its top. Similar growth mechanism is also seen during the formation of CACs of length $n\leq8$.}
\label{fig2}
\end{figure}

Single carbon ad-atom and mono-atomic chains of carbon atom are interesting entities, which can functionalize h-BN. A short segment of carbon atomic chain containing $n$ C atoms, indicated as CAC($n$) hereafter, is a truly one-dimensional carbon allotrope, and has recently drawn attention due to its linear geometry, high strength, size-dependent quantum ballistic conductance and interesting electronic properties. These properties of CACs were both theoretically\cite{abdurahman2002, tongay2004, tongay2005, senger2005, senger20051, dag2005, durgun2006, durgun20061, cahangirov2010, topsakal2010} and experimentally\cite{eisler2005, jin2009, chalifoux2010} investigated. Theoretical studies revealed the Peierls distortion in CACs,\cite{abdurahman2002, tongay2005, cahangirov2010} half-metallic and spintronic properties,\cite{dag2005, durgun2006, durgun20061} size dependent quantum conductance\cite{tongay2004, senger2005} and other geometrical structures.\cite{tongay2005} Chain structures of other group IV elements and group III-V compounds were also treated.\cite{senger20051} Concomitantly, carbon atomic chains are synthesized.\cite{eisler2005, jin2009, chalifoux2010} Recently, it was also shown that CACs can be grown on graphene and modify its properties,\cite{ataca2011} which have also been justified by the images obtained earlier using high resolution TEM.\cite{zettl2008}

In this paper we studied the growth of short carbon and BN chains on single layer h-BN and graphene, respectively. We first examined the adsorption of single carbon atom on h-BN by calculating its energy landscape and diffusion barrier. This is followed by the investigation of the nucleation and growth processes of CACs on h-BN. We performed both conjugate gradient calculations and molecular dynamics simulations in order to determine the stabilities and bonding properties of these CACs and show how they can grow on the plane of BN as new carbon atoms are introduced by one atom at a time at the close proximity of an existing CAC. Once these two materials, namely carbon atomic chains and monolayer h-BN, are combined fundamentally interesting properties are attained for promising future applications. Interestingly, the properties of the grown structures depend on the number of carbon atoms in the chains, such that they exhibit an even/odd disparity. In addition, we showed that CACs grown on h-BN constitute chemically active sites for Au, Li and H$_2$, which are normally weakly bonded to h-BN. For example, H$_2$ approaching to the top of the CAC(2) is dissociated and separated into two H atoms, each attached to the chain. We also showed that CACs can attach to another h-BN from its free end to serve as pillars to increase the spacing between h-BN flakes, where molecules like H$_2$ can be stored. We presented the electronic energy band structure for various lengths of chains to see their variations with the number of chain atoms. We finally studied the problem from the reverted point of view and presented the growth patterns, optimized structural and electronic structures of BN chains grown on graphene surface. Our results revealed the unique self-assembly character of carbon and BN chains on single layer honeycomb structures and interesting features attained thereof.

\section{Method}
In our calculations we use the state-of-the-art first-principles plane-wave calculations within the density-functional theory\cite{kohn1965,payne1992} combined with ab-initio, finite temperature molecular-dynamics calculations using projector augmented wave potentials.\cite{blochl1994,kresse1999} The exchange correlation potential is approximated by the generalized gradient approximation with Van der Waals correction.\cite{perdew1992,grimme2006} Numerical computations have been carried out by using VASP software.\cite{VASP} A plane-wave basis set with energy cutoff of $600 eV$ is used. The Brillouin zone is sampled in the \textbf{k}-space within the Monkhorst-Pack scheme,\cite{monkhorst1976} and the convergence of the total energy and magnetic moments with respect to the number of \textbf{k}-points are tested. The convergence for energy is chosen as $10^{-5}eV$ between two consecutive steps. In the relaxation of structures and band structure calculations, the smearing value for all structures is taken as $0.01eV$. We consider adsorption of chains on $(4\times4)$ supercells and treat the system using periodic boundary conditions. The pressure on each system was kept smaller than $\sim2kBar$ per unit cell in the calculations. In the ab-initio MD calculations the time step was taken as $2.5fs$ and atomic velocities were renormalized to the temperature set at $T=500K$ and $T=1000K$ at every $40$ time steps. In the MD stability tests, the simulations were run for $10ps$.

\section{Adsorption of single carbon ad-atom on h-BN}

Before going into detailed studies of carbon chains, we first investigate the adsorption and migration of single carbon ad-atom, which is the starting point of CAC growth on h-BN substrate. We use a $(4\times4)$ supercell of h-BN that consists of $16$ boron and $16$ nitrogen atoms. There is an energy difference of $\sim0.2eV$ between the spin polarized and spin unpolarized energy values in favor of the spin polarized state, indicating that the system has a magnetic ground state. Therefore all of the calculations mentioned hereafter are performed using the spin polarized conditions.

The most favorable binding site of single carbon atom was determined by placing the ad-atom initially to various adsorption sites at a height of $\sim2$\AA~ from the BN atomic plane and running fully self-consistent geometry optimization calculations by keeping the ad-atom fixed in $x-$ and $y-$ directions and letting the vertical $z-$coordinate of the ad-atom, which is its height from the plane, free. Meanwhile, the atoms in the BN supercell are relaxed in all directions except for one corner atom of the supercell, which is fixed in all directions to prevent h-BN from sliding.

In Fig.~\ref{fig1}(a) the nitrogen and boron atoms of the optimized h-BN structure are separated from each other by $1.45$\AA. The most favorable bonding site of single carbon ad-atom, which turns out to be near the top site of nitrogen atom, is marked with a red star. The migration(diffusion) path of carbon ad-atom on h-BN with a minimum energy barrier is shown by black stars. The minimum energy barrier is calculated as $0.68 eV$.\cite{canbn} The energy landscape calculated over the whole BN hexagon also shows that the energy barrier to the diffusion further increases as the carbon atom moves away from the nitrogen atom as shown in Fig.~\ref{fig1}(b). The variation of energy calculated along the 2D path shown in Fig.~\ref{fig1}(a) is presented in  Fig.~\ref{fig1}(c). As indicated in the figure, the most favorable site for the carbon atom is near the top nitrogen site, although not exactly on top of nitrogen. The energy barrier between the most favorable site and the top bridge(BR), top boron(TB), hollow(H) and top nitrogen(TN) sites were calculated as $0.07eV$, $0.95eV$, $1.00eV$ and $0.03eV$,  respectively.

In addition to the diffusion path analysis of a single carbon ad-atom, we next study the interaction between two carbon atoms on h-BN surface. When the distance between the two ad-atoms becomes less than a threshold distance of $\sim2$\AA, these two carbon atoms attract each other and form a CAC(2) perpendicularly attached to h-BN. This is indeed the nucleation for longer CACs. The most favorable binding site CAC(2), is again near the top site of nitrogen. A complete site analysis was also performed to confirm this result, by placing a CAC(2) on various adsorption sites and comparing the total energy values.

\section{Growth of carbon atomic chains on h-BN}

\subsection{Chain growth and even/odd disparity}

\begin{figure}
\includegraphics [width=8.5cm] {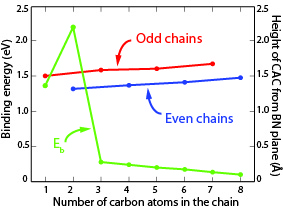}
\caption{Binding energies ($E_b$), and the heights($h$) of odd and even numbered CAC($n$)'s from the atomic plane of BN are shown in green, red and blue lines, respectively. The $h$ values exhibit an even/odd family behavior depending on the number of carbon atoms in the chain. The sudden peak in the binding energy arises from the change of the magnetic state of CAC(2) from magnetic to nonmagnetic when it binds to hexagonal BN.}
\label{fig3}
\end{figure}

Growth of the CAC further continues when a third carbon ad-atom is introduced at the close proximity of CAC(2). However, this time a complete site analysis of CAC(3) shows that the most favorable bonding site is the top of boron atom, instead of the nitrogen site. The formation of CAC(3) happens as follow: CAC(2) leaves its initial bonding position, moves higher from the BN plane and in the mean time it gets closer to the single ad-atom until they are bound to each other near the new energetic site, which is the top of boron site. Similar chain growth behavior is also seen during formation of CACs at different lengths. This process is further investigated with ab-initio molecular dynamics simulations at 500K and the snapshots taken from the growth of CAC(4) is presented in Fig.~\ref{fig2}. We initiate the MD simulation by placing a carbon ad-atom and a CAC(3) to their bonding sites as shown in Fig.~\ref{fig2}(i). The simulation was run for 2000 time steps and snapshots taken from the initial, 20$^{th}$, 40$^{th}$, and 120$^{th}$ time steps are shown. As the simulation proceeds to the 2000$^{th}$ step, the chain stays in its position shown Fig.~\ref{fig2}(iv), which is an indication of its stability at that bonding site.

\begin{table}
\caption{Most favorable binding sites, total binding energies($E_b$), magnetic moments($\mu$), heights($h$) of CAC($n$) from the BN plane, and the distances of the lowest carbon atom of the chain from the nitrogen($d_{C-N}$) and the boron($d_{C-B}$) atom in the BN plane for different $n$ values of carbon chains. The bonding sites and magnetic properties of CACs on BN exhibit an even/odd disparity. With the exception of the single carbon ad-atom, even numbered CACs bind to BN near the top of nitrogen(TN) atom and the odd numbered CACs bind near the top of boron(TB) atom. Additionally, the even and odd numbered chains grown on BN have magnetic and nonmagnetic(NM) ground states, respectively, with the exception of CAC(1) and CAC(2) cases. All calculations were performed on a $4\times4$ supercell which contains 32 carbon atoms.}
\label{table: 1}
\begin{center}
\begin{tabular}{ccccccc}
\hline  \hline
Structure & Site & $E_b(eV)$ & $\mu$($\mu_B$) & $h$(\AA) & $d_{C-N}$(\AA) & $d_{C-B}$(\AA) \\
\hline
BN+C & $\sim$TN & 1.36 & 2.00 & 1.50 & 1.59 & 1.73 \\
\hline
BN+CAC(2) & $\sim$TN & 2.19 & NM & 1.32 & 1.46 & 1.83 \\
\hline
BN+CAC(3) & $\sim$TB & 0.28 & NM & 1.59 & 2.16 & 1.62 \\
\hline
BN+CAC(4) & $\sim$TN & 0.24 & 2.00 & 1.37 & 1.50 & 1.63 \\
\hline
BN+CAC(5) & $\sim$TB & 0.20 & NM & 1.60 & 2.28 & 1.62 \\
\hline
BN+CAC(6) & $\sim$TN & 0.18 & 1.97 & 1.41 & 1.52 & 1.58 \\
\hline
BN+CAC(7) & $\sim$TB & 0.14 & NM & 1.67 & 2.41 & 1.64 \\
\hline
BN+CAC(8) & $\sim$TN & 0.08 & 1.96 & 1.48 & 1.56 & 1.59 \\

\hline
\hline
\end{tabular}
\end{center}
\end{table}

\begin{figure*}
\includegraphics [width=16cm] {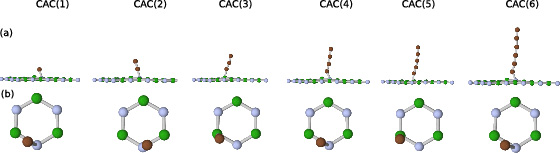}
\caption{Side and top views of the most favorable binding configurations of CAC($n$) on hexagonal BN are shown in (a) and (b). N, B, and C atoms are represented by blue, green and brown balls, respectively. CAC($n$)'s having an even number of carbon atoms (even $n$) is bound to BN near the top of nitrogen atom, whereas CAC($n$)'s with odd number of carbon atoms (odd $n$) prefer top boron site, with the exception of single carbon ad-atom. The geometries are calculated for a $(4\times4)$ supercell and their stabilities are also tested with MD simulations at $T=500K$ for $10ps$. In (b), only the carbon atom of the chain that is closest to the BN plane is shown.}

\label{fig4}
\end{figure*}

We further continue the above analysis with fourth, fifth, sixth and seventh carbon ad-atoms. In general, once a new carbon atom is implemented at the close proximity of the existing chain CAC($n-1$), this chain leaves its binding site and is bound to the top of new carbon ad-atom in their new most favorable binding site. Namely, this binding site keeps changing between the near top nitrogen site and the near top boron site. With the exception of the single carbon ad-atom, we observe an even/odd disparity of the binding site depending on the number of atoms in the CAC, that is even numbered CACs bind to h-BN substrate near the top nitrogen site, and the odd numbered CACs bind near the top boron site. This situation can be attributed to the charge density of CAC(n) at its end, which depends whether $n$ is even or odd.

The binding energies of CACs attached to graphene are calculated using the expression $E_{b}=E_{T}$[free-linear CAC(n)]+$E_{T}$[bare h-BN]-$E_{T}$[CAC(n)+h-BN]. In this expression, $E_{T}$[free-linear CAC(n)] is the total energy of fully relaxed linear chain of length $n$, $E_{T}$[bare h-BN] is the total energy of bare $(4x4)$ supercell of h-BN and $E_{T}$[CAC(n)+h-BN] is total energy obtained when CAC(n) binds to the supercell of h-BN. The complete list of these binding energies and the binding geometries of these CACs are presented in Table \ref{table: 1}.

It is seen that, with the exceptional case of CAC(2), which is further discussed in the following section, the binding energies($E_b$) of the CACs tend to decrease as the length of the chains (or the number of carbon atoms) increase. Similarly, due to the decrease in the binding energy, the height($h$), namely the distance between h-BN plane and the first carbon atom of the CAC, increases. However, the $h$ values also show an even/odd disparity. In other words, there is an increasing trend in the $h$ values when the CACs are grouped as even and odd numbered chains, but the height of CAC($2n$) is always less than CAC($2n-1$), as shown in Fig.~\ref{fig3}.  This situation can be explained by the fact thats the C-N bond length is shorter than the C-B bond length, and CAC(2n) is a more stable structure since it has a more symmetrical charge distribution as compared to CAC($2n-1$). These decrease $h$ distances of the even numbered CACs that are bound to the nitrogen atom.

We finally test the stabilities of these h-BN+CAC($n$) complexes at finite temperatures with molecular dynamics simulations. The ab-initio MD calculations were carried out for $10ps$ at $T=500K$. The final stable bonding configurations of various CACs are shown in Fig.~\ref{fig4}. After MD simulations, these CACs on h-BN remain stable and they don't detach from the BN plane, but they slightly pull the nearest B and N atoms of h-BN upwards from the plane, as seen from the side views given in  Fig.~\ref{fig4}(a). Also, the chains are not perfectly perpendicular to the BN plane, but are slightly tilted at different angles. Fig.~\ref{fig4}(b) shows the top views of the CACs, where only the carbon atoms closest to the BN planes are shown. As noted above, the even/odd disparity in the bonding sites of the chains with the exception of single carbon atom can be seen. It should be noted that ab-initio MD simulations carried out for 10ps is not sufficiently long to accumulate enough statistics. However, calculations at high temperatures are worthwhile to give a hint about the stability at relatively lower temperatures. An unstable chain would easily breakdown or detach from substrate. This did not happen in this case. Additionally, calculations using conjugate gradient method resulted in chain structures bound to the substrate with a significant binding energies, especially for short CACs. Hence, these tests provide sufficient evidence that the chains we studied are stable and remain attached to the single layer h-BN.

\subsection{Electronic and magnetic properties}

Having found the structural properties of CACs on h-BN, we next focus on the variations of magnetic and electronic properties of h-BN with various lengths of CACs attached. First, we perform both spin polarized and spin unpolarized energy minimization calculations for free(bare) CACs without attaching them to h-BN and compare the minimum energies of magnetic and non-magnetic cases. In a similar manner with the optimized structures of CACs attached to h-BN, it turns out the magnetization of free CACs also depends on the number $n$ of carbon atoms. Namely that a free chain has magnetic ground state if it has even number of carbon atoms, and a non-magnetic ground state if it has odd number of carbon atoms with the exception of single carbon atom in vacuum. Such an even/odd disparity in the magnetic moments of finite size CACs was previously reported.\cite{cahangirov2010} Hence, when these chains are adsorbed to h-BN sheet, the overall magnetic properties of the h-BN+CAC($n$) structures are maintained with the exception of CAC(2). While all other CACs preserve their magnetic ground states when bonded to h-BN, CAC(2) flips from the magnetic ground state to the non-magnetic ground state. This change in the magnetic moment of CAC(2) increases its bond strength with h-BN and causes a higher binding energy. This is the reason for the sudden jump in Fig.~\ref{fig3}.

\begin{figure}
\includegraphics [width=8.5cm] {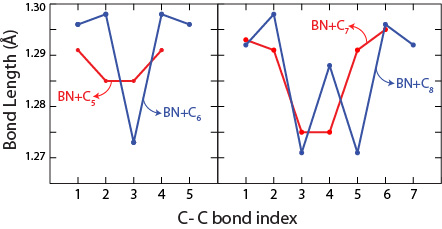}
\caption{Variation of C-C bond lengths in the short carbon atomic chains CAC($n$) grown on BN. While the even numbered chains (blue) have alternating short and long bonds, the odd chains(red) have a symmetrical bond length variation around their central atom. Note that the number of bonds in a chain is $n-1$}
\label{fig5}
\end{figure}

\begin{figure*}
\includegraphics[height=10cm]{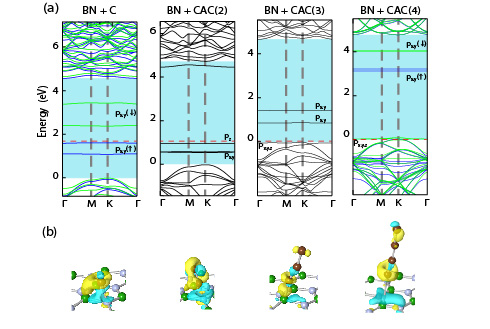}
\caption{(a) Electronic energy band structures of CACs grown on h-BN calculated for $n$= 1, 2, 3 and 4. In the magnetic cases, spin up and spin down bands are represented by blue and green lines, respectively. The bands below the red dash-dotted lines are fully occupied. The localized impurity states appearing as flat bands in the band gap originate from the $p$ bands of the carbon atoms that are at the edges of chains. (b) Isosurfaces of the difference charge densities of chains where yellow and green regions designate charge accumulation and charge depletion, respectively. The isosurface values are taken as 0.01 electron/\AA$^3$ for C, C$_2$, C$_3$ and as 0.005 electron/\AA$^3$ for C$_4$.}
\label{fig6}
\end{figure*}

The disparities observed so far are closely related to the bond length variation of CACs with even and odd number of atoms.\cite{cahangirov2010} First, we note that the final stable structures of short CACs grown on h-BN have nonuniform bond length distribution. While the even numbered chains have polyyne type of structure with alternating short and long bonds, the odd numbered chains have symmetrical bond length variations around their center points. In other words, the bond length variations for the odd chains reach to zero at the center, whereas the variation of bond lengths continues in chains having even $n$. This situation is related to the geometrical linear symmetry of the chain as also observed in a recent study for bare chains.\cite{fan2009} For the odd chains, the geometrical center of the chain is the middle carbon atom since there are even number of bonds. On the other hand, even chains have odd number of bonds and the geometrical center is on the middle bond. Since these chains are bonded to h-BN substrate only from one side they are free to move in the other direction. Hence, the chains need to be symmetrical around the center, and symmetrically equivalent bonds must have the same length. As a result, the two equivalent bonds around the center atom of the odd chains have the same length. Variations of bond lengths of h-BN+CAC($n$) for $n$=5-8 are presented
in Fig.~\ref{fig5}

This overall symmetry in the geometrical structure of the odd numbered chains also leads to a symmetry in the distribution of the magnetic moments, making the ground states singlet for odd chains. In the singlet state, magnetic moments exerted by the electrons cancel each other leading to zero net spin and zero magnetic moment. Therefore, these structures have a nonmagnetic ground state. On the other hand, even chains have an uneven distribution of magnetic moments on atoms leading them to have magnetic ground states. This magnetic preference also alters the binding sites of the chains accordingly. One would expect this situation to be reverted for chains saturated from both ends because the boundary conditions for the chains are altered as shown in a recent study.\cite{cahangirov2010}

An individual CAC($n$) attached to h-BN can modify the electronic band structure of h-BN. If an adsorbed CAC($n$) is sufficiently far from others, it gives rise to localized states in the band gaps and resonance states in the band continua of h-BN. In this study we consider a single CAC($n$) adsorbed to h-BN using a model where one CAC($n$) is attached to each repeating ($4\times 4$) supercell with the condition that interactions between adjacent CACs are negligible. This model recovers approximately the ($4 \times 4$) folded bands of bare h-BN, except that the energy difference between the top of the valance bands and the bottom of the conduction bands gradually increases from $4.5eV$ to $4.8eV$. Additionally, flat bands due to CAC($n$) occurs in the band gap. These flat bands actually corresponds to the localized states due to CAC($n$). In Fig.~\ref{fig6}(a) we present the electronic energy structure of h-BN+CAC($n$) calculated for $n$=1,2,3 and 4 using supercell geometry explained above. The energy positions of the filled and empty flat bands in the gap are closely related with the energy levels of the corresponding CAC($n$), which vary with $n$. For magnetic $n$=1 case, spin-up states (i.e. flat bands) originating mainly from C-$p_{xy}$ orbitals occur above the top of valence band. Empty spin-down states of the same orbital character occur near mid gap. In the non-magnetic case of $n$=2 spin-degenerate C-$p_{xy}$ and C-$p_{z}$ states are filled and occur near the top of the valence band. While a filled C-$p_{xyz}$ state touches the top of valence band both for non-magnetic $n$=3 case and magnetic $n$=4 case, the positions of empty C-$p_{xy}$ states in the band gap strongly depend on $n$. For all $n$'s the resonance states occur in the valence band within the energy of 2eV from the top.

The difference charge density is calculated by subtracting the charge densities of bare h-BN and bare CAC($n$) from that CAC($n$)+h-BN system by keeping the atomic configuration of adsorbed and bare CAC unaltered. In Fig.~\ref{fig6}(b) the isosurfaces of difference charge density indicate that important changes from the corresponding bare CAC is found where CAC is bound to h-BN.

\subsection{Applications of CACs on BN}

\begin{figure*}
\includegraphics [width=18cm] {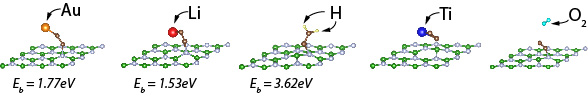}
\caption{Functionalization of BN sheets through adsorption of carbon chains. For example, a CAC(2), which is strongly bound to h-BN, creates chemically active sites for Au, Li and H atoms. H$_2$ molecule approaching to CAC(2) from sides dissociated to form two C-H bonds, whereas O$_2$ remains totally inactive. Ti atom takes the carbon atoms with itself and forms TiH$_2$. }
\label{fig7}
\end{figure*}

Apart from its interesting and fundamental aspects, the growth of CACs on single layer graphene is a reality and they can modify graphene's physical and chemical properties. Experimental observations and images obtained using high resolution TEM\cite{zettl2008} indicates that CACs can be formed on monolayer graphene surfaces. In this study, we show that BN is also a suitable substrate for the growth of CAC($n$). Previous experimental studies\cite{liu2011} and density functional results\cite{ozcelik2012, haghi2012} have shown that epitaxial graphene grown on boron nitride surface is possible by means of chemical vapor deposition technique or exposure of  BN layer to carbon atoms or dimers at long time intervals. In these studies, it was noted that defected graphene structures with undesired chains can also form during the growth of graphene.\cite{jin2009, topsakal2010} Thus, since the temperature range of our MD calculations are suitable for atomic layer deposition techniques, it is conjectured that similar  atomic layer deposition or molecular beam epitaxy methods can be used for chain growth on BN layers at specific ambient conditions.

\begin{figure}
\begin{center}
\vspace{1cm}
\includegraphics [width=8cm]{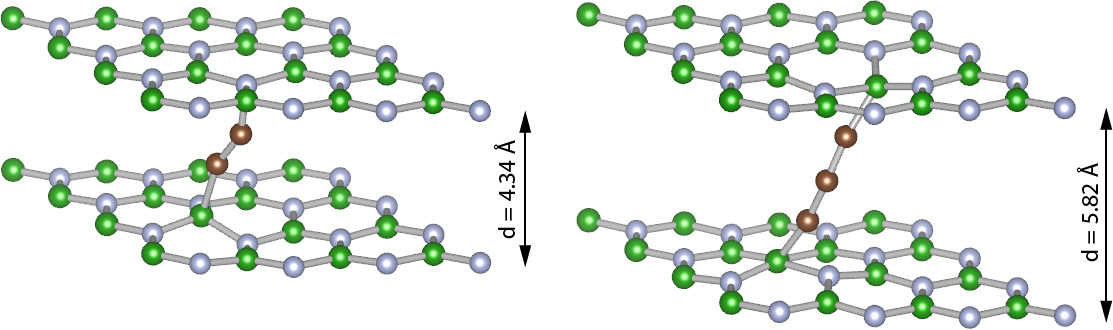}
\end{center}
\caption{CAC(2) and CAC(3) grown between two BN flakes. The optimized spacing between the flakes increase from $3.1$\AA~ to $4.34$\AA~ and $5.82$\AA~ upon the formation of chains.}
\label{fig8}
\end{figure}

BN is in general not a chemically active material, however with the inclusion of certain ad-atoms its chemical activity can be improved. CACs grown on BN can be used for functionalization of h-BN layers which will enable the adsorption of certain ad-atoms like hydrogen, lithium and gold. This functionalization will especially be useful where the contact on an insulating surface like BN is desired along with connection of electrodes. In order to increase the chemical activity of the inactive BN layer we use CAC(2) chain. In the previous sections, we showed that CAC(2) binds to h-BN with a very high binding energy as compared to other CACs. Besides being bonded strongly to h-BN, it also creates a chemically active sites for the absorption of other ad-atoms.

Here we test the adsorption of Au, Li, Ti, H$_2$ and O$_2$ on CAC(2) and show that the chemical activity of BN can be enhanced through CAC(2) attached to it. Ad-atoms were placed in the vicinity of CAC(2) on h-BN and fully self consistent calculations were performed. Before introducing the ad-atoms, the CAC(2) was placed freely in its most favorable position as calculated in the previous section. It was seen that Au and Li atoms move towards the chain and are bound to the chemically active site of CAC(2) with binding energies of $1.77$ and $1.53eV$, respectively. These are significantly higher than the binding energies of Au and Li on bare h-BN, which are calculated as $0.01$ and $0.12eV$, respectively. This is an important result indicating that gold electrodes can easily be connected to semiconducting h-BN through CACs. A hydrogen molecule approaching to CAC(2) from the sides moves upwards and dissociates to form two C-H bonds. The binding energy for H atoms is calculated as $3.62eV$. However, when O$_2$ is introduced to the system instead of H$_2$, it stays completely inactive and stay away from CAC(2). Finally, we consider Ti ad-atom. Normally, Ti is bound to bare h-BN with an energy of $0.80eV$. However, when h-BN is functionalized with CAC(2), although Ti ad-atom initially is bound to the carbon atoms, it doesn't stay there but takes away these CAC(s) from the BN layer and forms a TiC$_2$ structure which moves away from BN. This final configuration is energetically more favorable than Ti binding to h-BN by $5.9eV$. The final relaxed geometries of all these structures are presented in Fig.~\ref{fig7}.

Pillared graphene structures and CACs passivated by graphene surfaces from both ends have also been subjects of recent interest of both theoretical and experimental studies.\cite{dimitrakakis2008} It was recently shown that CACs passivated by graphene flakes from both ends produce highly stable chain structures.\cite{ravagnan2009} Oxidized graphene pillars were also shown to be useful materials for storage applications.\cite{burress2010}

Motivated by previous studies, we also calculated the bonding geometries of CACs between BN layers and showed that CACs can also be grown between two BN flakes as shown in Fig.~\ref{fig8}. We have demonstrated this situation by calculating optimized bonding configurations and the spacing between two BN flakes when CAC(2) and CAC(3) are grown between them. CAC(2) is bonded to nitrogen atom from one side and to boron atom on the other side. On the other hand, CAC(3) is bonded to both of the flakes from the top of boron atom. Once CAC(2) and CAC(3) are grown, the spacing between the flakes increases from $3.1$\AA~ to $4.34$\AA~ and $5.82$\AA, respectively. The formation energies, calculated by subtracting the energy of two BN planes and the energy of CAC from the energy of the final structure, are found as $2.0$ and $0.32eV$ for CAC(2) and CAC(3), respectively.

\begin{table}
\caption{Most favorable binding sites, binding energies($E_b$) and the heights($h$) of B, N ad-atoms and BN or NB atomic chains from the graphene plane. Here, for example, BN(3) indicates the BN chain consisting of three atoms (namely N-B-N) grown on graphene with N being attached to graphene.}
\label{table: 2}
\begin{center}
\begin{tabular}{cccc}
\hline  \hline
Structure & Binding site  &  $E_b(eV)$ & $h(\AA)$ \\
\hline
B-ad-atom & Bridge & 1.24 & 1.45 \\
\hline
N-ad-atom & Bridge & 1.05 & 1.25 \\
\hline
NB(2) & Bridge & 2.58 & 1.46 \\
\hline
BN(2) & Top  & 2.42 & 1.32 \\
\hline
NB(3) & Bridge & 1.18 & 1.52 \\
\hline
BN(3) & Top  & 0.90 & 1.31 \\
\hline
NB(4) & Bridge & 0.82 & 1.58 \\
\hline
BN(4) & Top  & 0.46 & 1.39 \\
\hline
NB(5)& Bridge & 0.73 & 1.63 \\
\hline
BN(5) & Top  & 0.21 & 1.48 \\
\hline
NB(6) & Bridge & 0.51 & 1.67 \\
\hline
BN(6) & Top  & 0.15 & 1.57 \\

\hline
\hline
\end{tabular}
\end{center}
\end{table}

\section{BN atomic chains grown on graphene}

\begin{figure}
\begin{center}
\includegraphics [width=8cm]{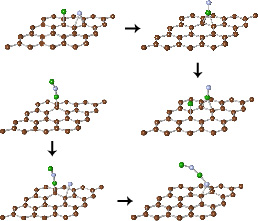}
\end{center}
\caption{Growth dynamics of short NB chains on graphene. Once a new ad-atom is placed at a random point in the neighborhood of a NB chain having $n$ atoms, the structure rearranges itself to form a chain comprising $n+1$ atoms. The snapshots are taken from the quantum MD simulations performed at 500K for 10ps. Blue, green and brown balls are, respectively, nitrogen, boron and carbon atoms.}
\label{fig9}
\end{figure}

Having shown the stable carbon chain structures grown on BN, we next investigate the growth of short BN chains on graphene to see if a similar self-assembly mechanism is also present for this reverted situation. Here we adopt the representation where BN($n$)+graphene indicates a BN chain consisting of $n$ atoms that is attached to graphene through N atom. For example, NB(3) + graphene (or NB(3) only) stands for B-N-B chain attached to graphene through B atom. Following the similar procedures described in Sections III and IV, we calculate most favorable binding sites, electronic properties, binding energies. We first perform MD simulations to see the self assembly mechanisms of BN chains. To this end we start by placing single B and N ad-atoms on graphene supercell and run MD simulations at 500K for 10ps. Eventually, the two ad-atoms form a BN molecule perpendicularly attached to graphene. When a third ad-atom N is placed on the graphene at the close proximity of existing BN molecule, MD simulations show that a chain comprising three atoms, B-N-B forms. This procedure was repeated for different lengths and it was observed that B and N atoms are indeed self-assembled on graphene honeycomb structure to form short BN atomic chains. Snapshots taken from these MD simulations are presented in Fig.~\ref{fig9}.

\begin{figure*}
\includegraphics [width=16cm]{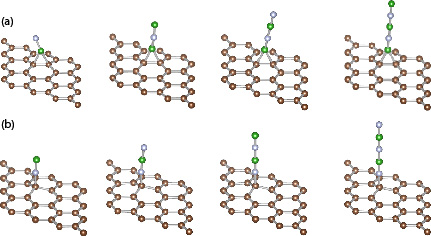}
\caption{The most favorable binding configurations of short NB and BN chains grown on graphene. A NB($n$) chain is attached to graphene at the bridge site, since its nearest atom to graphene is B, and BN($n$) is bound at the top of carbon atom, since its nearest atom to graphene is N. Energies are calculated using a (4x4) supercell and their stabilities are tested with MD simulations at T = $500K$ for $10$ ps. Note that the most favorable binding site is the bridge site for both single N and single B ad-atoms. Blue, green and brown balls are, respectively, nitrogen, boron and carbon atoms.}
\label{fig10}
\end{figure*}

\begin{figure}
\begin{center}
\includegraphics [width=8.5cm]{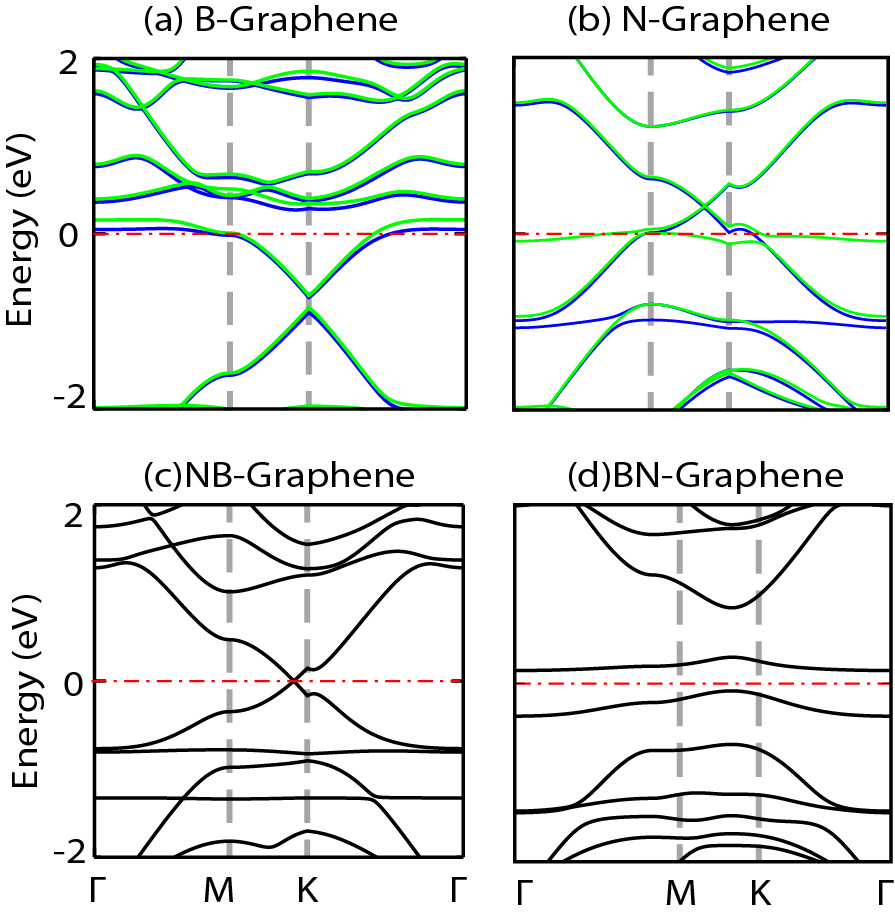}
\end{center}
\caption{Electronic energy structures: (a) Single B ad-atom adsorbed on graphene. (b) Single N ad-atom adsorbed on graphene. c) NB chain consisting of two atoms grown on graphene with B atom being closest to graphene. d) BN chain of two atoms grown on graphene with N being the closest ad-atom to graphene. The zero of energy is set to the Fermi energy, shown by red dash-dotted line. In the magnetic cases, spin up and spin down bands are represented by blue(dark) and green(light) lines, respectively.}
\label{fig11}
\end{figure}

The most favorable binding sites of short BN chains on graphene and their corresponding binding energies were calculated with self consistent conjugate gradient calculations. It was found that, the binding site of a BN chain on the graphene depends on the type of its closest atom to the graphene. A NB($n$) chain is bound to the graphene layer at the bridge site, and a BN($n$) is bound at the top site as shown in Fig.~\ref{fig10}. However, both single B and single N ad-atoms are bound at the bridge site. Contrary to the CAC case, no clear even/odd disparity was obtained in h-BN+graphene complexes owing to the ionic bonding between B-N atoms. Here there is another effect related with the number of electrons in the chain. Chains having odd number of atoms have odd number of electrons, while even numbered chains have even number of electrons. The binding energies of these chains shown in in Table \ref{table: 2} suggest that BN chains are bound with higher energies to the graphene as compared to the binding energies of CACs on BN, especially for longer chains. The binding energy of a boron ending chain (i.e. NB) is always higher than a nitrogen ending chain (i.e. BN) site. Here we note that the supercell size and adsorbate-adsorbate

The even/odd disparity observed in the magnetic properties of CAC($n$)+h-BN are also not seen for BN chains grown on graphene.  Single B ad-atom as well as single N ad-atom adsorbed on graphene have magnetic ground states. However, while BN(2) and NB(2) chains on graphene are nonmagnetic, BN(3) and NB(3) on graphene have magnetic ground states. This order is disturbed by the chain consisting of four atoms; while BN(4)+graphene is nonmagnetic, NB(4)+graphene has magnetic ground state.

The electronic structures calculated for a single adsorbate, namely single boron and nitrogen ad-atoms, and single NB(2) and BN(2) molecules, perpendicularly attached to (4x4) supercell of graphene are shown in Fig.~\ref{fig11}. Owing to their odd number of electrons, single N and B ad-atoms shift the Fermi level up and down relative to the Dirac point of graphene. This is an important feature to be used in electronic applications of single layer graphene. For NB(2)+graphene the bands crossing at the Fermi level attribute semimetallic character. On the other hand, BN(2)+graphene is a semiconductor with a narrow indirect band gap. This situation shows the dramatic effect of the type of chain atom which is attached to graphene. In addition, the supercell size, adsorbates-adsorbate coupling, as well as the symmetry of the adsorbate+supercell system are crucial for the resulting electronic structure for small supercells.\cite{hasanmesh,lambin} For example, the semiconductor presented in Fig.~\ref{fig11} (d) becomes metallic when BN+graphene system is treated in (6x6) supercell. In the case of very large supercell having single adsorbate attached the band crossing is recovered and the states of adsorbate appear as localized (resonance) impurity states.

\section{Conclusion}
In conclusion, using first-principles calculations within the density functional theory, we revealed the growth mechanism of carbon and BN atomic chains on single layer honeycomb structures, namely hexagonal BN(or h-BN) and graphene, respectively. We found that with the inclusion of each new carbon atom, the existing atomic chain consisting of $n$ carbon atom leaves its previous position to join with the new carbon atom to from a chain of $n+1$ carbon atom. Similar, but more complex growth mechanisms were also found for BN chains grown on graphene layers depending on the type of chain atom(B or N) which is attached to graphene. These growth processes were simulated by ab-initio molecular dynamics calculations at various temperatures and the resulting chain structures were shown to be stable even though they are free to bend and slightly tilt. Nevertheless, these simulations are performed at high temperatures(1000K) and the chains did not detach from the substrate plane for 10ps which is adequately long for ab-initio molecular dynamics simulations. Therefore these simulations presents reliable stability results, especially at room temperature. The growth of atomic chains on the single layer honeycomb structures heralds a self-assembly process, which may have fundamental and technological implications.

The grown chains by themselves, exhibit interesting physical and chemical properties depending on the number of atoms forming the chain and the type of chain atom attaching to the substrate. In particular, the physical properties of even numbered and odd numbered carbon chains can behave differently leading to interesting even-odd disparity. We also showed that atomic chains grown on single layer substrates attribute useful functionalities to bare h-BN and graphene. These properties are dependent on the number of carbon atoms in the chains.  Apart from creating localized electronic states in the band gap and local magnetic moments it is demonstrated that carbon chains can also be used for increasing the interlayer spacing between BN flakes, where specific molecules can be stored. Atomic chains grown on graphene and h-BN create chemically active sites on BN surface for atoms such as Au, Li and H$_2$, which provides the connection of these materials. Additionally, specific atoms forming strong bonds with atomic chains modify the electronic structure in the band gap and hence change the conductivity for possible sensor applications. Similarly, modifications of the electronic properties of graphene depending on the grown B or N ad-atoms and BN or NB chains are also worth emphasizing. Moving the Fermi level of graphene up or down by adsorbing N or B atoms, respectively can be a worthwhile feature in the electronic applications related with graphene. The self-assembly in terms of atomic chains grown on semimetallic graphene and semiconducting h-BN and interesting functionalities achieved thereof can bring up new perspectives for further research on single layer honeycomb structures.

\section{Acknowledgements}
Part of the computational resources has been provided by TUBITAK ULAKBIM, High Performance and Grid Computing Center (TR-Grid e-Infrastructure) and UYBHM at Istanbul Technical University through Grant No. 2-024-2007. This work was supported partially by the Academy of Sciences of Turkey(TUBA).

\end{document}